\documentstyle [12pt] {article}
\title{A SIMPLE THEORETICAL PREDICTION OF THE DATA CORRESPONDING TO OBSERVATIONALLY
ESTIMATED VALUE OF COSMOLOGICAL CONSTANT }

\author{Vladan Pankovi\'c$^{\ast,\sharp}$,Simo
Ciganovi\'c$^\sharp$,\\
Jovan Ivanovi\'c$^\sharp$, Rade Glavatovi\'c$^\diamond$, Petar Gruji\'c$^\Box$ \\
$^\ast$Department of Physics, Faculty of Sciences, 21000 Novi
Sad,\\ Trg Dositeja Obradovi\'ca 4. , Serbia, vdpan@neobee.net \\
$^\sharp$Gimnazija, 22320 Indjija, Trg Slobode 2a, Serbia\\
$^\Diamond$ Military-Medical Academy, 11000 Belgrade, Crnotravska
17., Serbia\\
$^\Box$ Institute of Physics, 11 000 Belgrade, P. O. Box 57,
Serbia, grujic@phy.bg.ac.yu \\ }

\date {}
\begin{document}
\maketitle

\vspace {0.5cm}
\vspace {0.5cm}

\begin {abstract}
In this work a satisfactory, simple theoretical prediction of the
data corresponding to observationally (by fine tuning condition)
estimated value of the cosmological constant is given. It is
supposed (in conceptually analogy with holographic principle) that
cosmological constant, like classical surface tension coefficient
by a liquid drop, does not correspond to a volume (bulk) vacuum
mass (energy) density distribution but that it corresponds to a
surface vacuum mass (energy) density distribution. Then form of
given surface mass distribution and fine tuning condition imply
observed growing (for $\sim$ 61 magnitude order) of the scale
factor (from initial, corresponding to Planck length, to recent,
at the beginning of the cosmic acceleration, corresponding to 10
Glyr length).
\end {abstract}
\vspace {0.5cm}

As it is well-known (see for example [1]), Friedmann equation
without cosmological constant term
\begin {equation}
   (\frac{1}{a} \frac {da}{dt})^{2} + \frac {kc^{2}}{{\it a}^{2}}= G\frac {8\pi}{3}\rho
\end {equation}
(where  ${\it a}$ represents the  scale factor of the universe,
$k$ - curvature constant (that equals 1 for closed, 0 for flat and
-1 for open universe), $c$ - speed of light, $G$ - Newtonian
gravitational constant, $\rho$ - total mass density) can be
formally interpreted by classical, Newtonian mechanics and
gravitation in the following way. Namely, equation (1) can be
equivalently, simply transformed in the following equation
\begin {equation}
   \frac{1}{2}(m \frac {dr}{dt})^{2} + \frac {kmc^{2}}{2}= m G\frac {4\pi}{3} r^{3}\frac {\rho}{r}
\end {equation}
where $r=a r_{0}$ represents sphere radius, $r_{0}$ - appropriate
length unit, while $m$ can be considered as the mass of a
classical probe system, a particle or spherical shell. Term
$\frac{1}{2}(m \frac{dr}{dt})^{2}$ can be formally considered as
the classical kinetic energy. Constant term $\frac {kmc^{2}}{2}$
can be formally considered as the total energy of a classical
harmonic linear oscillator or rotator. Finally, term $ mG(\frac
{4\pi}{3})r^{3}\frac {\rho}{r} $ can be formally considered as the
classical potential energy by gravitational interaction between
probe system and universe with mass $(\frac {4\pi}{3})r^{3}\rho$
homogeneously distributed by density $\rho$ within the sphere
volume $(\frac {4\pi}{3})r^{3}$.

Friedmann equation with additional cosmological constant term
\begin {equation}
   (\frac{1}{a}\frac {da}{dt})^{2} + \frac {kc^{2}}{{\it a}^{2}}=
   G\frac {8\pi}{3}\rho + \frac {\Lambda}{3}
\end {equation}
(where $\Lambda$ represents the cosmological constant) can be,
also, formally classically interpreted. Namely, it can be
equivalently, simply transformed in the following equation
\begin {equation}
   \frac{1}{2}(m \frac {dr}{dt})^{2} + \frac {kmc^{2}}{2}=
   m G\frac {4\pi}{3} r^{3}\frac {\rho}{r} + m\frac {\Lambda}{24 \pi} (4\pi r^{2})
\end {equation}
Additional term $ m(\frac {\Lambda}{24 \pi}) (4\pi r^{2}) $ can be
formally considered as the classical energy of the surface tension
of a fluid captured in the sphere with radius $r$ and surface area
$(4\pi r^{2}) $ so that surface tension coefficient is $ m(\frac
{\Lambda}{24 \pi}) $. It is very important to be pointed out that
here (as well as in the fluid mechanics), even if surface tension
coefficient is constant, energy of the surface tension increases
when sphere radius increases.

All this implies a possibility [2] that cosmological constant, not
only formally classically, but even really corresponds to some
surface phenomena. (Especially, it can be observed that all this
corresponds, at least conceptually, to t'Hooft-Susskind
holographic principle in quantum gravity [3].) One aspect of this
possibility we shall roughly consider in this work. Namely, in
this work a satisfactory, simplified theoretical prediction of the
data corresponding to observationally (by fine tuning condition)
estimated value of the cosmological constant [4], [5] will be
given. It will be supposed that cosmological constant, like
surface tension coefficient by a liquid drop, does not correspond
to a volume (bulk) vacuum mass (energy) density distribution but
that it corresponds to a surface vacuum mass (energy) density
distribution. Precisely, it will be supposed that vacuum mass is
distributed over a thin spherical shell with sphere radius
proportional to scale factor and thickness equivalent to Planck
length. Then form of given surface mass distribution and fine
tuning condition imply observed value of the scale factor, i.e.
growing of the scale factor (for $\sim 61$ magnitude order) from
initial (corresponding to Planck length $\sim 10^{-35}m$) to
recent (at the beginning of the cosmic acceleration, corresponding
to $10 Glyr \sim 10^{26}m$ length).

So, suppose, usually,
\begin {equation}
   G\frac {8\pi}{3} \rho_{\Lambda} = \frac {\Lambda}{3}
\end {equation}
where $\rho_{\Lambda}$ represents the mass density corresponding
to cosmological constant.

Further, suppose
\begin {equation}
  \rho_{\Lambda} = \frac {M_{\Lambda}}{L_{P}4\pi r^{2}}
\end {equation}
which corresponds to vacuum mass $ M_{\Lambda}$ homogeneously
distributed over thin spherical shell with sphere radius $r$ and
thickness equivalent to Planck length $ L_{P}$. It implies
\begin {equation}
  M_{\Lambda}  = \rho_{\Lambda}( L_{P}4\pi) r^{2}
\end {equation}
which means that vacuum mass grows quadratically when sphere
radius grows.

Introduction of (6) and
\begin {equation}
  \Lambda = \frac {c^{2}}{L^{2}_{\Lambda}}
\end {equation}
(where $L_{\Lambda}$ represents the length corresponding to
cosmological constant) in (5), after simple transformations,
yields
\begin {equation}
  \frac {M_{\Lambda}}{M_{P}}  = \frac {1}{2} \frac {r^{2}}{ L^{2}_{\Lambda}}
\end {equation}
It represents very interesting result admitable for comparison
with observational data.

Initially, i.e. for $r = L_{P}$, and, according to fine-tuning
condition [4], [5]
\begin {equation}
     \frac { L^{2}_{P}}{ L^{2}_{\Lambda}} \sim 10^{- 123}
\end {equation}
it follows
\begin {equation}
  \frac {M_{\Lambda}}{M_{P}}= \frac {1}{2} \frac { L^{2}_{P}}{ L^{2}_{\Lambda}}
\end {equation}
or
\begin {equation}
  M_{\Lambda} = 10^{- 123} M_{P}
\end {equation}
It means, of course, that cosmological constant does not any
important influence in the early universe.

Now, we shall determine by (9) such $r$ for which condition
\begin {equation}
    M_{\Lambda}\sim M_{P}
\end {equation}
is satisfied. Given condition, of course, simply means that vacuum
energy becomes comparable with energy of the quantum fields.
Introduction of (13) in (9) yields, after simple transformations,
\begin {equation}
   r^{2} \sim  L^{2}_{\Lambda}
\end {equation}
Now, we shall express $r$ in the following way
\begin {equation}
   r = {\it a} L_{P}
\end {equation}
which, introduced in (14), yields
\begin {equation}
  {\it a}^{2} \sim \frac { L^{2}_{\Lambda}}{ L^{2}_{P}} \sim 10^{123}
\end {equation}
and further
\begin {equation}
  {\it a} \sim 10^{61}
\end {equation}
It represents very important result. Namely, it satisfactorily
corresponds to observational data [4]. [5] on the growing of the
scale factor of the universe for $\sim61$ magnitude order ,from
initial (corresponding to Planck length $\sim 10^{-35}m$) to
recent (at the beginning of the cosmic acceleration, corresponding
to $\sim 10 Glyr \sim 10^{26}m$ length). (It can be added that, as
it is not hard to see, supposition on the volume, i.e. bulk
distribution of the vacuum mass yields unsatisfactory prediction
${\it a} \sim 10^{41}$).

Finally, we can shortly repeated and pointed out the following. It
can be pointed out that suggested model of the interpretation of
cosmological constant as a surface phenomena (at least
conceptually analogous to t'Hooft-Susskind holographic principle)
is very rough and simplified, formally like to a classical. Such
model must be necessarily generalized, but it goes over basic
intention of our work. In any case our model is able to correlate
and reproduce observational astronomical data (fine-tuning and
growing of the scale factor) in a satisfactory way. All this is
very interesting and promising.

\section {References}

\begin {itemize}

\item [[1]] B. W. Carroll, D. A. Ostlie, {\it An Introduction to Modern Galactic Astrophysics and
Cosmology}
(Addison-Wesley, Reading, MA, 2007)
\item [[2]] T. Padmanabhan, {\it Dark Energy: Mystery of the Millennium}, astro-ph/0603114 and references therein
\item [[3]] R. Bousso, {\it The holographic principle}, hep-th/0203101 and references therein
\item [[4]] D. N. Spergel et al., Astrophys. J. Supp. {\bf 146} (2003) 175 ; astro-ph/0302209.
\item [[5]] D. N. Spergel et al., Astrophys. J. Supp. {\bf 170} (2007) 337 ; astro-ph/0603449.

\end {itemize}

\end {document}